\documentclass[a4paper,doc,natbib]{apa6}
\usepackage[english]{babel}
\usepackage[utf8x]{inputenc}
\usepackage{amsmath}
\usepackage{amssymb}
\usepackage{graphicx}
\usepackage{pdflscape}
\usepackage[colorinlistoftodos]{todonotes}
\usepackage{mathtools}
\usepackage{algorithm}
\usepackage{url}
\newcounter{mntcomm}
\newcommand{\wt}{\widetilde}

\def\N{{\cal N}}

\title{Time-Evolving Psychological Processes Over Repeated Decisions}

\shorttitle{Modeling Time-on-Task} 

\author{David Gunawan$^{1,5}$, Guy E. Hawkins$^2$, Robert Kohn$^{3,5}$, Minh-Ngoc Tran$^4$, Scott D. Brown$^2$}

\affiliation{1: School of Mathematics and Applied Statistics, University of Wollongong \\
 2: School of Psychology, University of Newcastle, Australia \\
 3: Australian School of Business, University of New South Wales, Sydney, Australia \\
 4: Discipline of Business Analytics, University of Sydney Business School\\
 5: Australian Research Council Center of Excellence in Mathematical and Statistical Frontiers}

\authornote{This research was partially supported by the Australian Research Council (ARC) Discovery Project scheme (DP180102195, DP180103613). Hawkins was partially supported by an ARC Discovery Early Career Researcher Award (DECRA; DE170100177). Earlier versions of this work are posted online at \url{https://arxiv.org/abs/1906.10838} and presented at the 2019 meetings of the Psychonomic Society and the Australasian Mathematical Psychology Conference. We thank the editor and reviewers for detailed and constructive feedback. Their engagement improved this manuscript. Correspondence concerning this work should be addressed to scott.brown@newcastle.edu.au}


\abstract{Many psychological experiments have subjects repeat a task to gain the statistical precision required to test quantitative theories of psychological performance. In such experiments, time-on-task can have sizable effects on performance, changing the psychological processes under investigation. Most research has either ignored these changes, treating the underlying process as static, or sacrificed some psychological content of the models for statistical simplicity. We use particle Markov chain Monte-Carlo methods to study psychologically plausible time-varying changes in model parameters. Using data from three highly-cited experiments we find strong evidence in favor of a hidden Markov switching process as an explanation of time-varying effects. This embodies the psychological assumption of ``regime switching'', with subjects alternating between different cognitive states representing different modes of decision-making. The switching model explains key long- and short-term dynamic effects in the data. The central idea of our approach can be applied quite generally to quantitative psychological theories, beyond the models and data sets that we investigate.
}

\begin{document}
\maketitle

Almost all experiments in cognitive psychology involve repetition: subjects make repeated responses to very similar (or even identical) stimuli for an extended period of time. Repeating trials reduces measurement noise, and trades off against the number of subjects from whom data are collected. In paradigms with little variation between subjects (e.g., some aspects of vision), many thousands of observations are collected from just a handful of people. In other experiments where large variation is expected between people (e.g., social attitudes), a small number of data are collected from many hundreds of subjects.

A typical experiment is the investigation of perceptual decision-making reported by \cite{forstmann2008striatum}, where subjects made several hundred decisions about the direction of apparent motion in a random dot stimulus. On some trials, subjects were encouraged to respond slowly and carefully, on other trials urgently, and on other trials to balance caution and urgency (these conditions are addressed later). Figure~\ref{fig:datatrends} shows the data differently from the many other analyses reported elsewhere for this experiment. The basic properties of subjects' decisions change substantially over the course of the experiment. The left panels of Figure~\ref{fig:datatrends} show that the average accuracy of decisions decreases with time-on-task, while both the average response time (RT) and its variability increases. The data for individual subjects is even more variable; the right panels of the figure show the data for three subjects. Different individuals certainly show different-sized effects of time-on-task, but they also appear to have different qualitative directions of the effect. For example, the top right panel of the figure shows that average accuracy is approximately stable for one subject, decreases with time-on-task for another, and increases with time-on-task for the third. Changes with practice are to be expected, because psychological experiments are both difficult and taxing. Subjects learn to be better at the task, but they also suffer from fatigue, boredom, and many other unpredictable effects. It is unreasonable to assume that identical cognitive processes underlie each repeated decision.

We advance the investigation of time-on-task effects by combining recent advances in statistical estimation with modern cognitive modeling methods. The key idea is to account for dynamic changes by allowing the parameters of a cognitive model to change with time-on-task according to flexible but statistically tractable dynamic models. Our approach combines advantages of previous methods, such as coherently pooling information across different times, while avoiding some of their limitations, such as assuming independence between times, or assuming rigid functional forms for the time effects. In addition to providing a more sophisticated way to address the effects of time-on-task, our approach allows the investigation of scientifically meaningful models of dynamics and sequential effects, embedded within scientifically meaningful models of the cognitive process. An important outcome of our new approach is the ability to distinguish between competing explanations for dynamic effects based on very different underlying assumptions. For example, the analyses below compare theories based on smooth changes in underlying processes over time (a``trend'' model) from non-smooth changes in the process (an autoregressive model) and also from a stationary process with different hidden states (a ``regime switching'' model).


\subsection{Previous Approaches}
\begin{figure}
\centering
\includegraphics[width=0.6\textwidth]{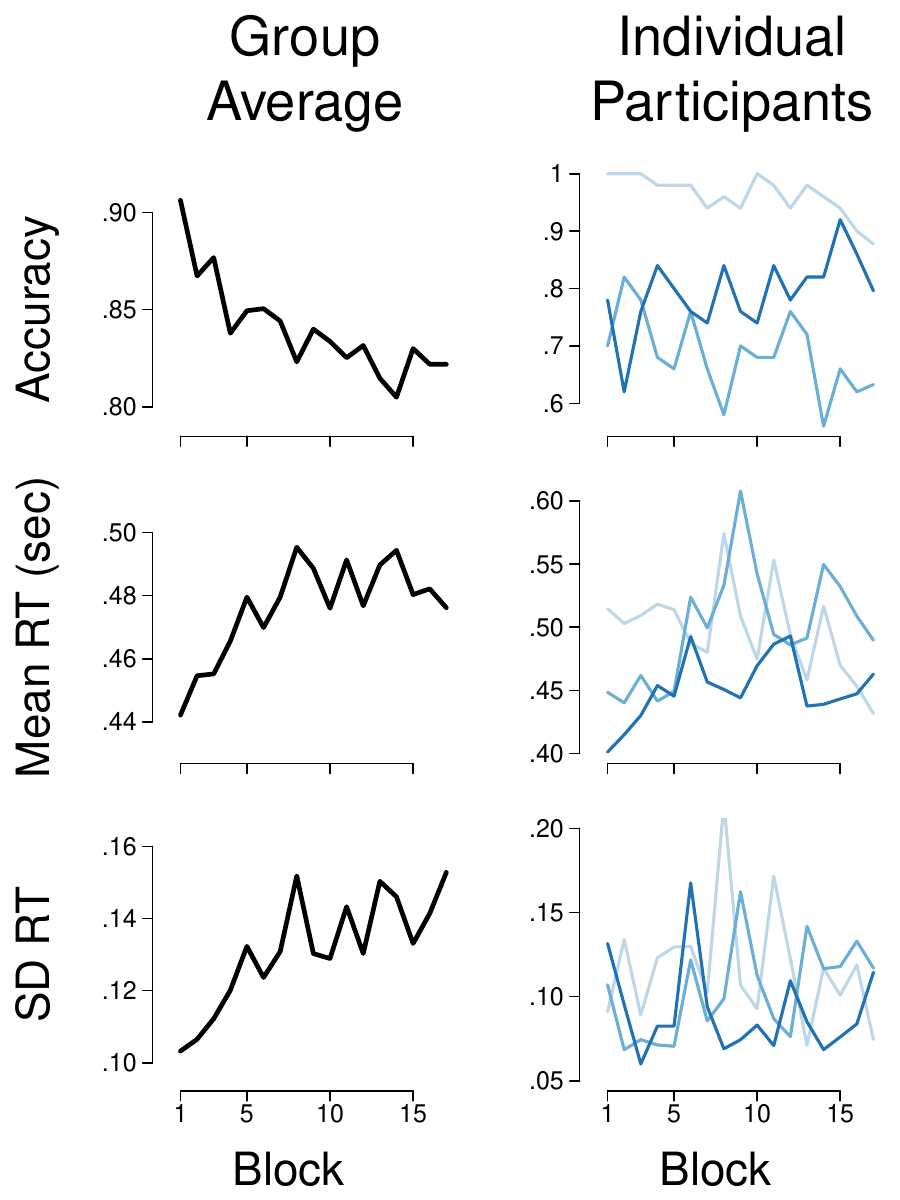}
\caption{\label{fig:datatrends} Data from \cite{forstmann2008striatum}. Subjects make repeated decisions about simple stimuli, but the effects of practice are complex and varied. The top row shows that accuracy in those decisions decreases with time-on-task ($x$-axis) in the aggregate (left panels), but this trend is inconsistent across three example subjects (separate lines, right panels). The middle panels show that mean response time (RT) across subjects increases with practice, but the same three individual subjects show a mix of increasing or decreasing RT with practice. Similarly complex trends are apparent in the variance of RT (bottom panels).}
\end{figure}

Performance changes with time-on-task are often sufficiently large that ignoring them is unreasonable. In the perceptual decision-making experiment described above, average accuracy of decisions declines from about 90\% at the start of the experiment to just above 80\% at the end (Figure~\ref{fig:datatrends}). This effect is just as large as the key experimental manipulation studied by \cite{forstmann2008striatum}; the differences between cautious, urgent, and neutral decision-making. The average RT and accuracy are still changing by the end of the session,  even after substantial practice. Some experiments include even more practice than that -- we investigate two examples of this later. In some of these cases we observe stable group summary statistics, like mean accuracy and RT, after sufficient practice. However, even this does not imply that any individual’s performance is constant. This is the same problem associated with averaging that arises in most analyses. Our investigations (see Figure~\ref{fig:goodness-of-fit-RT-LDT1}, for example) suggest that most subjects continue to change substantially on either, or both, of mean RT and accuracy even up to 2,000 trials of practice.

There have been efforts to understand the psychological causes of these changes, through study of the effects of extended practice \citep{newell1981mechanisms,palmeri1999theories,heathcote2000power,evans2018refining,kahana2018variability}, of fatigue \citep{ratcliff2011diffusion,walsh2017computational}, or ``mind wandering'' \citep{christoff2016mind-wandering,mittner2016neural}. Those researchers sought to understand changes over time by applying time-varying quantitative models to the data. Some of these efforts were based on psychological process models \citep[e.g.,][]{palmeri1999theories,christoff2016mind-wandering} and others on descriptive statistical models \citep[e.g.,][]{kahana2018variability}. Both approaches have strengths, but progress has been hampered because the quantitative modeling of time-varying data presents challenging computational and statistical problems. 
This is evident, for example, in the difficulty of addressing important and widely-recognized questions about the moment-to-moment ``sequential effects'' in behavioral data. Sequential effects refer to the micro-structure of dynamic effects, where subjects' responses are strongly influenced by their experiences and behavior in the preceding few seconds \citep[e.g.,][]{kim2017bayesian,craigmile2010autocorrelated,brown2008integrated}. Traditional approaches to modeling dynamic effects either require strong assumptions prescribing the way in which behavior changes over time, or require splitting the data into smaller and smaller segments, bringing the usual problems associated with small samples.

Examples help illustrate the limitations of existing approaches to modeling time-varying effects. \cite{newell1981mechanisms}, \cite{heathcote2000power} and \cite{evans2018refining} investigated how the performance of simple skills (e.g., rolling cigars, memorizing rules, ...) improves with practice. The researchers took a ``strong assumptions'' approach, by identifying a few candidate functions that might describe how performance changes with practice, and comparing those functions against the data. This approach is tractable and powerful, but there are severe limits imposed by the particular functions that are chosen for comparison. Their investigations focus primarily on the comparison of just two theories about how performance changes with practice (exponential vs. power functions). The approach is difficult to extend to very many comparison functions. The chosen functions also represent strong theoretical assumptions about changes with practice, and about uniformity across people and time.


As examples of the other type of approach mentioned above, \cite{evans2019humans} and \cite{kahana2018variability} both investigated performance under conditions of extended practice, in decision-making and memory recall paradigms respectively. The effects of time-on-task were addressed by splitting the data from each subject into periods defined by the design (different sessions, from different days, and different blocks within sessions). Using blocks and sessions as units of analysis allowed the researchers to tease apart effects due to the stimuli, the subjects, and practice. This approach is powerful, but depends upon very substantial data sets. In both experiments, each subject completed many hours of participation, over days or weeks. For more typical experiments, splitting data to investigate time-on-task can lead to unacceptably small sample sizes.

\section{A New Approach}

We illustrate and test our new theories of task dynamics using data from three experiments: the one discussed above \citep{forstmann2008striatum} and two reported by \cite{wagenmakers2008diffusion}. We adopt a hierarchical Bayesian approach to the analysis, with each subject having a unique set of parameters for a cognitive model of the task, and constrain these parameters to follow a parametric distribution across subjects. We explore three different accounts of dynamic change and the micro-structure of decision-making. For two accounts, we allow the parameters of the model to change across time in the experiment, according to simple statistical processes: either smooth polynomial trends, or flexible non-smooth changes described by an autoregressive process. For the third account, we investigate a model motivated by a process-level psychological theory of regime switching. This account encapsulates the idea that subjects can make decisions using different modes or ``states'', such as on-task and off-task, or more vs. less cautiously. The model explains how subjects switch between those different states, using a hidden Markov process. The state-based hidden Markov process is motivated by ``regime switching'' accounts of mind-wandering and decision-making \citep[e.g.][]{mcvay2010does,mittner2016neural,smallwood2015science,kahneman2011thinking}, which propose binary cognitive changes with unpredictable switches between hidden states.

The dynamic processes provide constraints on the evolution of parameters across subjects, but they do not impose fixed deterministic values for changes with time. This facilitates exploration of the differences between people in changes with time, which are expressed in the posterior estimates. For example, the polynomial trend prior captures the expectation that average trends are similar across people. The autoregressive prior imposes the expectation that the ``persistence'' (correlation) of parameters from block to block is the same across people. The Markov switching process imposes the expectation that there is a probability of being ``off-task'' for each person, but does not force this to be identical across people, or across time-on-task. We explore the practical and scientific performance of these assumptions in psychological data. This requires developing new estimation approaches that are based on particle MCMC methods. While our examples are particular to the data and models we explore, an important contribution of our work is that the statistical approach and the new estimation methods are quite general and likely to be useful in a wide range of investigations. 

\section{Methods}

\subsection{Modeling Approach\label{subsec:Time-varying LBA model}}

\begin{figure*}
\begin{center}
\includegraphics[keepaspectratio,width=14cm]{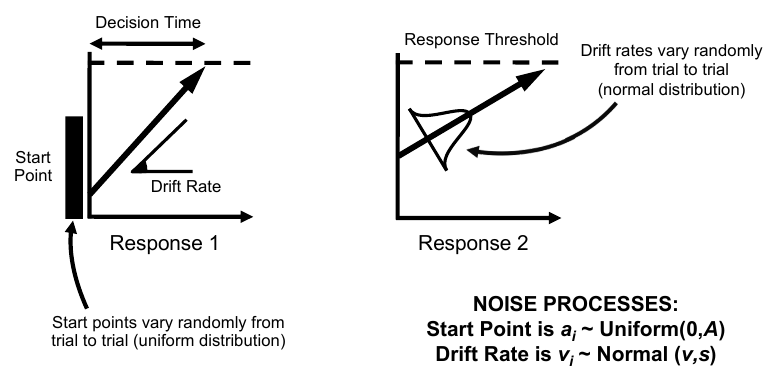}
\end{center}
\caption{The linear ballistic accumulator (LBA) model of decision-making represents a choice as a race between two accumulators. The accumulators gather evidence until one of them reaches a threshold, triggering a decision response. The time taken to make the decision is the time taken to reach the threshold, plus a constant offset amount for processes unrelated to the decision. Variability arises from the starting point of the evidence accumulation process and its speed; both vary randomly from trial-to-trial and independently in each accumulator. }
\label{fig:lba}
\end{figure*}

All three data sets considered here consist of repeated simple decisions. It is standard to model such data using evidence accumulation models \citep{ratcliff2016diffusion,donkin2018response}. We use the linear ballistic accumulator \citep[LBA;][]{brown2008simplest}, which is a well-established accumulator model that was previously applied to data including those analyzed here. Figure~\ref{fig:lba} illustrates that the LBA represents a decision between two options (such as ``word'' vs. ``non-word'' or ``leftward motion'' vs. ``rightward motion'') as a race between two accumulators. Each accumulator gathers evidence in favor of one of the two responses. The amount of evidence in each accumulator increases with passing decision time, until the evidence in one of the accumulators reaches a threshold amount, triggering a decision response. The model makes quantitative predictions about the joint distribution over response times and decision outcomes. These predictions are specified by the values given to the model's parameters: the height of the response threshold, the distributions of drift rates and starting points for evidence accumulation, and the amount of time taken by processes outside of the decision itself.

\subsubsection{Standard (Static) LBA Model} Modern applications of the standard static LBA model use a hierarchical Bayesian implementation. Mostly, these applications use differential evolution Markov chain Monte-Carlo (DE-MCMC) to estimate the posterior distribution over parameters and random effects \citep{turner2013method}. Our analyses use a more efficient estimation procedure based on particle MCMC methods developed by \cite{gunawan2020new}.

Decisions in the experiments under consideration are always forced choices between two alternatives. For these, the $i^{th}$ decision for the $j^{th}$ subject yields two pieces of information. The first is the response choice, $RE_{ij}\in\left\{ 1,2\right\}$. The second is the response time, $RT_{ij}\in\left(0,\infty\right)$. Assuming that the variance of the drift rate distributions in the LBA model are always $s^2=1$ \citep[see also][]{donkin2009overconstraint}, the predictions of the model for any particular decision are defined by a vector of five parameters; $\left\{ \wt b_{j} ,\wt A_{j}, \wt v_{j}^{\left(1\right)}, \wt v_{j}^{\left(2\right)},\wt \tau_{j} \right\}$. Parameter $\wt b_{j}$ is the decision threshold adopted by subject $j$ -- the amount of evidence required to trigger a decision. The tilde notation (such as $\wt b_{j}$) is used to distinguish these parameters, which are restricted to positive values, from the log-transformed ones we work with (such as $b_{j}$). Parameter $\wt A_{j}$ is the width of the uniform distribution of start points for the same subject, and $\wt \tau_{j}$ is the amount of time taken by processes other than evidence accumulation (also called the ``non-decision time''). The values $\wt v_{j}^{\left(1\right)}$ and  $\wt v_{j}^{\left(2\right)}$ give the mean of the drift rate distributions for the two racing accumulators. The model parameters are constrained sensibly by the experimental conditions. For example, the thresholds are allowed to be different for speed-emphasis vs. accuracy-emphasis trials, but are constrained to be equal for all speed-emphasis trials. \cite{gunawan2020new} develop efficient methods for estimating this model, allowing for parameters which vary from person to person as random effects. The random effects are assumed to be normally distributed, after log-transformation. 
Computer code to use the model, based on Gunawan et al.'s algorithms and functions from \cite{terry2015generalising}, are freely available as R packages, ``PMwG'' \citep{pmwg} and ``rtdists'' \citep{rtdists}.

\subsubsection{Three Time-Varying LBA Accounts} Three extensions of the standard (static) LBA model that allow time-varying parameters for individual subjects are considered. We call these the ``AR'', ``trend'', and ``switching'' models. The AR and trend models assume that the parameters for any subject in any given block of trials are constant, but that those parameters evolve over blocks. The two models differ in the prior (group-level distribution) that they assume for this evolution. The AR model assumes a first-order autoregressive prior and the trend model assumes a polynomial regression prior with linear and quadratic trends. 


For the AR and trend models, we examine the effects of time-on-task by splitting the data into blocks representing different time periods. While coarser and finer splits are possible in principle, we respect the designs of the original experiments by using blocks defined by their procedures. Using blocks to define time-on-task (instead of, e.g., single trials) is an important choice because it also allows random effects for blocks -- different parameter values in different blocks, for each person, but with these constrained to follow person-level distributions across blocks and group-level distributions across people. This approach is similar to the multi-level modeling (or ``linear mixed effects'') taken in other situations, and used by \cite{kahana2018variability} to investigate session-by-session variability in memory performance. 

For the {\bf autoregressive (AR)} process we assume a simple ``AR1'' model, in which the parameters governing performance in each block are random evolutions of the parameters from just the previous block (i.e., lag=1). For each subject, $j=1,...,S$, and for each block, $t=1,...,T$, 
the density of log-transformed random effects for subject $j$ in block $t$, $\alpha_{j,t}$, is
\begin{eqnarray}
\alpha_{j,t}|\alpha_{j,t-1},\phi,\mu,\Sigma & \sim & \N\left(\mu+\textrm{diag}\left(\phi\right)\left(\alpha_{j,t-1}-\mu \right),\Sigma\right),\label{eq:TV-LBA(I)}\\
\alpha_{j,1}|\mu,\Sigma & \sim & \N\left(\mu,
\Sigma\right),\label{eq:TV-LBA(I)_2}
\end{eqnarray}
where $\textrm{diag}\left(\phi\right)$ is the diagonal matrix with the diagonal vector of autoregressive coefficients $\phi=\left(\phi_1,...,\phi_D\right)$; $D$ is the dimension of $\alpha_{j,t}$, and for parsimony $\phi$ is assumed to be the same for all subjects ($j$) and its elements are constrained to the interval $(-1,1)$ to ensure that distant time blocks are less correlated. 
Values of $\phi$ near to 1 indicate a high correlation between adjacent blocks, while values of $\phi$ near zero mean that the blocks are almost independent of each other. Equation~\eqref{eq:TV-LBA(I)_2} initializes the $\alpha_{j,t}$ sequence by providing a prior distribution for the random effects in the first time period, $\alpha_{j,1}$; the location of this prior distribution is $\mu$ and its covariance matrix is $\Sigma$, that is, the group-level multivariate normal distribution.


Our definition for the AR prior has useful properties for a psychological account of time-on-task. For example, it implies that the expected value (in the prior) of the random effects do not change with time. The expected correlations between random effects are approximately constant across blocks, and the correlations between the same random effect in different blocks falls away exponentially with the distance between the blocks. Exploring other autoregressive models is possible within the framework we have developed; for example including longer effects of stimulus and response history.

We assume the following priors for the group parameters in Equations \eqref{eq:TV-LBA(I)} and \eqref{eq:TV-LBA(I)_2}: $\mu \sim \N\left(0,I_{D}\right)\;\;\text{and}\;\;
\Sigma \sim IW\left(v=20,S_{\alpha}=I_{D}\right)\label{eq:priorforsigma}$; $IW$ denotes the inverse Wishart distribution, which is a standard prior assumption for covariance matrices, and is conjugate with the multivariate normal distribution; see e.g., \cite{gelman2014bayesian-data-analysis}. For the vector of autoregressive paqrameters The prior for each of the autoregressive parameters $\phi_i, i=1, \dots, D$ is uniform on $(-1,1)$  For the vector of autoregressive parameters, we assumed independent uniform priors, $\phi_i \sim U\left(-1,1\right)$ for $i=1,...,D$.

For the {\bf polynomial trend} model, we assumed a regression model with linear and quadratic effects of time-on-task (block) on the parameters. For each subject $j=1,...,S$, and for each time block, $t=1,...,T$, the density of the log-scaled random effect $\alpha_{j,t}$ is
\begin{eqnarray}
\alpha_{j,t}|\mu_{t},\Sigma & \sim & \N\left(\mu_{t},\Sigma\right);\label{eq:TV-LBA(II)}
\end{eqnarray}
the $d^{th}$ component of $\mu_{t}$ is $\beta_{d1}+\beta_{d2}t+\beta_{d3}t^2$, $d=1,...,D$.
The polynomial coefficients have a standard normal distribution: $\beta_{dk}\stackrel{i.i.d.}{\sim} N\left(0,1\right)$, $d=1,...,D$, $k=1,2,3$ and the prior for $\mu$ and $\Sigma$ is the same as in the AR model.


The time-varying {\bf Markov switching} model is different from the other two in that it does not operate on blocks of trials, but at the level of individual trials. It assumes that each person's decision-making is governed by two separate vectors of random effects, representing two different mental states or two different modes or even ``systems'' for decision-making. These random effects are estimated from data and can differ in a wide variety of ways. For example, they could differ in the sensitivity of the decision-making process (drift rates), which might correspond to on-task vs. off-task states. Alternatively, they could differ in the speed-accuracy tradeoff parameters (thresholds or start points), which correspond to different strategies for balancing decision speed vs. caution. 


The trial-by-trial specification of the Markov switching model, compared to block-by-block for the AR and trend models, provides an opportunity to capture the micro-structure of decision-making. The switching model assumes that each subject has two different states in which they might perform the task,\footnote{We limit our modeling to $K=2$ states, but note that the method extends naturally to any number of hidden states.} defined by different parameter settings for the evidence accumulation process. 
The subject performs the decision-making task on any given trial in either one or the other of these states. The hidden Markov model describes the process by which the subject switches between states. On each trial, there is a fixed probability of switching from the current state to the other state. This probability can be different for switching in the different directions (i.e., a different probability for switching out of state 1 into state 2 than for switching out of state 2 into state 1), which allows the Markov transition matrix to be asymmetric. Psychologically, this is important as it allows for a non-uniform steady state distribution over states with one state more likely than the other.

It is psychologically plausible that the random effects for the two states are correlated -- a subject who is faster than average in one state is also likely to be faster than average in the other state. To support this, we estimated a model in which the prior allowed for correlations in the random effects between states. Each subject's performance was described by a vector of random effects, $\alpha_j$, which was twice the dimension as that for the other models ($2\times D$), formed by concatenating the random effect vectors for the two states. For each subject, we also estimated two random effects for the switching probabilities. For subject $j$, we denote these by $\xi_{12}$ and $\xi_{21}$, for the probabilities of switching out of states 1 and 2, respectively. Full details of the sampling algorithm and other estimation methods for the Markov switching LBA model are available in the online supplementary material: \url{http://osf.io/x29wb}.

\subsubsection{Model Estimation} 

The statistical difficulties associated with reliably estimating time-varying cognitive models have previously presented barriers to the theoretical progress with time-varying effects, including of practice, fatigue, and learning. Our methods build on recent developments in statistical treatments for the static LBA model \citep{gunawan2020new}. Our new estimation approaches use particle Metropolis within Gibbs (PMwG) to estimate the time-varying random effects models by defining an augmented parameter space which includes copies of all the model's parameters, and the trajectories (history) of the particles representing these. This enables Metropolis-within-Gibbs sampling while guaranteeing that the samples are generated from the exact posterior.

We ran the PMwG algorithm with the number of particles fixed at $R=250$. Three different sampling stages are employed; in the initial stage, the first $1,000$ iterates are discarded as burn-in; the next $4,000$ iterates are used in the adaptation stage to construct the efficient proposal densities for the random effects for the final sampling stage; finally, a total of $10,000$ MCMC posterior draws are obtained in the sampling stage. For inference about the models' performance, we extended the importance sampling based method of \citet{tran2021robustly}, called $\textrm{IS}^{2}$. This robustly and efficiently estimates the marginal likelihood of each dynamic model, thus supporting inference via Bayes factors.   We ran the $\textrm{IS}^{2}$ algorithm with $M=10,000$ importance samples to estimate the log of the marginal likelihood. The number of particles used to obtain the unbiased estimate of the likelihood was set to $R=500$, and the Monte Carlo standard errors of the log of the marginal likelihood estimate were obtained by bootstrapping the importance samples. The details of the estimation methods are given in online supplements which include statistical properties related to our new algorithms and practical computational details. The code to implement all the algorithms is available from the same site (\url{http://osf.io/x29wb}), along with material which demonstrates the statistical robustness of the methods, including a simulation study showing that our methods appropriately recover the data-generating process and data-generating parameters of the models.

\section{Data}


\subsection{Dynamic Effects in Perceptual Decision-Making} \cite{forstmann2008striatum} has 19 subjects make repeated decisions about the direction of motion shown in a random dot kinematogram. We analyzed data from the behavioral training, pre-scanning, session in which each subject made 840 decisions distributed evenly over three conditions. These conditions changed the instructions given to subjects about whether they should emphasize the speed of their decisions, the accuracy of their decisions, or adopt a ``neutral'' balance between speed-emphasis and accuracy-emphasis. The data reveal large changes in both the speed (response time) and accuracy of decisions between speed-emphasis and accuracy-emphasis conditions, but only small differences between the accuracy-emphasis and neutral-emphasis conditions. Full details of the method are on p.17541 of the original article.

To model the decisions in these data, we follow the same LBA specification as in the original article. 
The modeling collapses across left- and right-moving stimuli, forcing the same mean drift rate for the accumulator corresponding to a ``right'' response to a right-moving stimulus as for the accumulator corresponding to a ``left'' response to a left-moving stimulus; let $v^{\left(c\right)}$ be this mean drift rate. Similarly, drift rates for the accumulators corresponding to the wrong direction of motion are constrained to be equal and denoted by $v^{\left(e\right)}$. Three different response thresholds are estimated, for the speed, neutral, and accuracy conditions: $B^{\left(s\right)}$, $B^{\left(n\right)}$, and $B^{\left(a\right)}$ respectively. Two other parameters are estimated: the time taken by non-decision process ($\tau$) and the width of the uniform distribution for start points in evidence accumulation ($A$). For each emphasis condition $k$, we estimate $\wt B^{\left(k\right)} = \wt b^{\left(k\right)} - \wt A$ instead of directly estimating $ \wt b^{\left(k\right)}$, ensuring that the response threshold is always larger than the maximum value of the starting point of evidence accumulation.

To investigate time-on-task for the AR and polynomial trend models, we divided the trials into blocks matching the experimental procedure, of size $n=50$ trials on average (there is some variability due to a few missing data). There are $T=17$ blocks for each subject. The random effects in the AR and polynomial trend model evolve over blocks, which we denote by subscript $t$. For the Markov switching model, the random effects do not change with blocks; dynamic effects instead arise from switching between the two internal states. These states are represented by different values for the random effects, which means there are twice as many as for the other two models.

\subsection{Dynamic Effects in Lexical Decision-Making -- Block-by-Block Speed Accuracy Tradeoff}

The subjects in Experiment 1 of \cite{wagenmakers2008diffusion} made decisions about whether letter strings are valid English words (e.g., ``RACE'') or non-words (e.g., ``RAXE''). Each of 17 subjects made decisions about $1,920$ letter strings. Half of the letter strings are non-words. The other half are divided across three types of words: high frequency words which are very common in written English (e.g., ``ROAD''); low frequency words which are uncommon (e.g., ``RITE''); and very low frequency words, which are extremely rare (e.g., ``RAME'').  Decisions are arranged into $T=20$ blocks of $n=96$ decisions (trials) each. The instructions given to subjects change from block-to-block; in alternate blocks, subjects are instructed to emphasize the accuracy or the speed of their decisions. Full details of the method are on p.144 of the original article. We use the experimenter-defined blocks to investigate the effects of time-on-task for the AR and trend models: $n=96$ trials in each of $T=20$ blocks, except for a very small number of missing trials for some subjects.

To describe the lexical decisions with the LBA model, we assume that: (a) the speed-emphasis vs. accuracy-emphasis manipulation influences only threshold settings; and (b) the different stimulus categories (word frequency) influences only the means of the drift rate distributions. These are sometimes called ``selective influence'' assumptions, and are important for the psychological interpretation of the theory \citep{ratcliff1998modeling,voss2004interpreting}. For the AR and polynomial trend models, our assumptions result in the following random effects for subject $j$ in block $t$,

\begin{equation}
\begin{split}
\big({B_{j,t}^{\left(s\right)}},{B_{j,t}^{\left(a\right)}},{A_{j,t}},{v_{j,t}^{\left(hf,W\right)}},{v_{j,t}^{\left(lf,W\right)}},{v_{j,t}^{\left(vlf,W\right)}},{v_{j,t}^{\left(nw,W\right)}},{v_{j,t}^{\left(hf,NW\right)}},{v_{j,t}^{\left(lf,NW\right)}},\\
{v_{j,t}^{\left(vlf,NW\right)}},{v_{j,t}^{\left(nw,NW\right)}},{\tau_{j,t}}\big),\label{eq:raneff-ldt1}
\end{split}
\end{equation}

The random effects include independent mean drift rates ($v$) for each stimulus class (\emph{hf, lf, vlf, nw}) and response (\emph{W, NW}). The Markov switching model assumes the same structure, but with a vector of random effects which is twice as long. This captures different random effects for the two internal states, and does not assume changes over blocks in the nature of the states.

\subsection{Dynamic Effects in Lexical Decision-Making -- Effects of Base Rate} Experiment 2 of \cite{wagenmakers2008diffusion} is very similar to Experiment 1, but uses 19 new subjects. The $1,920$ trials are split into alternating blocks of 96 trials each. The blocks are either dominated by words or non-words: in some blocks there are 24 words and 72 non-words, and in the other blocks there are 72 words and 24 non-words. Blocks alternate between non-word-dominated and word-dominated. Before each block begins, subjects are reminded which kind of stimulus string will be most common in the next block of trials. Full details of the method are on p.152 of the original article. We again use the experimenter-defined blocks to investigate the effects of time-on-task ($n=96$ and $T=20$).

To describe the lexical decisions in this experiment, we allow for response bias, expressed as different decision thresholds in the accumulators corresponding to ``word'' ($B^{\left(.,W\right)}$) and ``non-word''  ($B^{\left(.,NW\right)}$) responses. These are allowed to be different between the blocks dominated by word ($B^{\left(w,.\right)}$) and non-word ($B^{\left(nw,.\right)}$) stimuli, following the hypothesis that the base rate of the stimulus classes should influence subjects' biases. For the AR and polynomial trend models, our assumptions result in the following random effects for subject $j$ in block $t$,

\begin{multline}
\left({B_{j,t}^{\left(w,W\right)}},{B_{j,t}^{\left(w,NW\right)}},{B_{j,t}^{\left(nw,W\right)}},{B_{j,t}^{\left(nw,NW\right)}},{A_{j,t}},{v_{j,t}^{\left(hf,W\right)}},{v_{j,t}^{\left(lf,W\right)}},{v_{j,t}^{\left(vlf,W\right)}},\right.\\
\left.{v_{j,t}^{\left(nw,W\right)}},{v_{j,t}^{\left(hf,NW\right)}},{v_{j,t}^{\left(lf,NW\right)}},{v_{j,t}^{\left(vlf,NW\right)}},{v_{j,t}^{\left(nw,NW\right)}},{\tau_{j,t}}\right).\label{eq:raneff-ldt2}
\end{multline}

Again, the Markov switching model uses a vector of random effects which is twice as long, for the two different internal states.

\section{Results}

\subsection{Model Performance}

Table~\ref{tab:BFdata} shows that, for all three experiments, the Markov switching model provides the best explanation of the data. The differences between the marginal likelihoods for the models are large compared to the scales usually used for such comparisons; e.g., they correspond to Bayes factors much larger than $10^6$ for all pairwise comparisons. The Monte-Carlo errors of the log of the estimated marginal likelihoods (in brackets) are small, confirming the practical efficiency of the method. For all experiments, the autoregressive (AR) model has the second-best marginal likelihood. The time-varying models provide a much better explanation of the data than the standard static LBA. The static model provides a poorer account than all three time-varying models in almost every case; the only exception is that the static model outperforms the polynomial trend model in the data from \cite{forstmann2008striatum}.

\begin{table}
\caption{Selecting between the static LBA model and the three time-varying versions, using data from three experiments. The log of the marginal likelihood estimates show that the preferred model for all three experiments is the Markov switching model. The standard errors are in brackets.}

\centering{}%
\begin{tabular}{c|cccc}
& \emph{Static} & \emph{AR} & \emph{Trend} & \emph{Switching} \tabularnewline
\hline 
\hline
Forstmann et al. (2008) & $\underset{\left(0.05\right)}{7459}$ & $\underset{\left(0.12\right)}{{7889}}$ & $\underset{\left(0.16\right)}{7417}$  & $\underset{\left(3.31\right)}{\textbf{8199}}$ \tabularnewline 
Wagenmakers et al. (2008) Exp. 1 & $\underset{\left(0.17\right)}{5970}$ & $\underset{\left(0.18\right)}{{8184}}$ & $\underset{\left(0.53\right)}{8003}$  & $\underset{\left(2.86\right)}{\textbf{8305}}$ \tabularnewline 
Wagenmakers et al. (2008) Exp. 2 & $\underset{\left(0.35\right)}{9630}$ & $\underset{\left(0.75\right)}{{10,324}}$ & $\underset{\left(0.46\right)}{9915}$  & $\underset{\left(3.86\right)}{\textbf{11,292}}$ \tabularnewline
\end{tabular}
\label{tab:BFdata}
\end{table}

\begin{figure}
\centering
\includegraphics[width=0.85\textwidth]{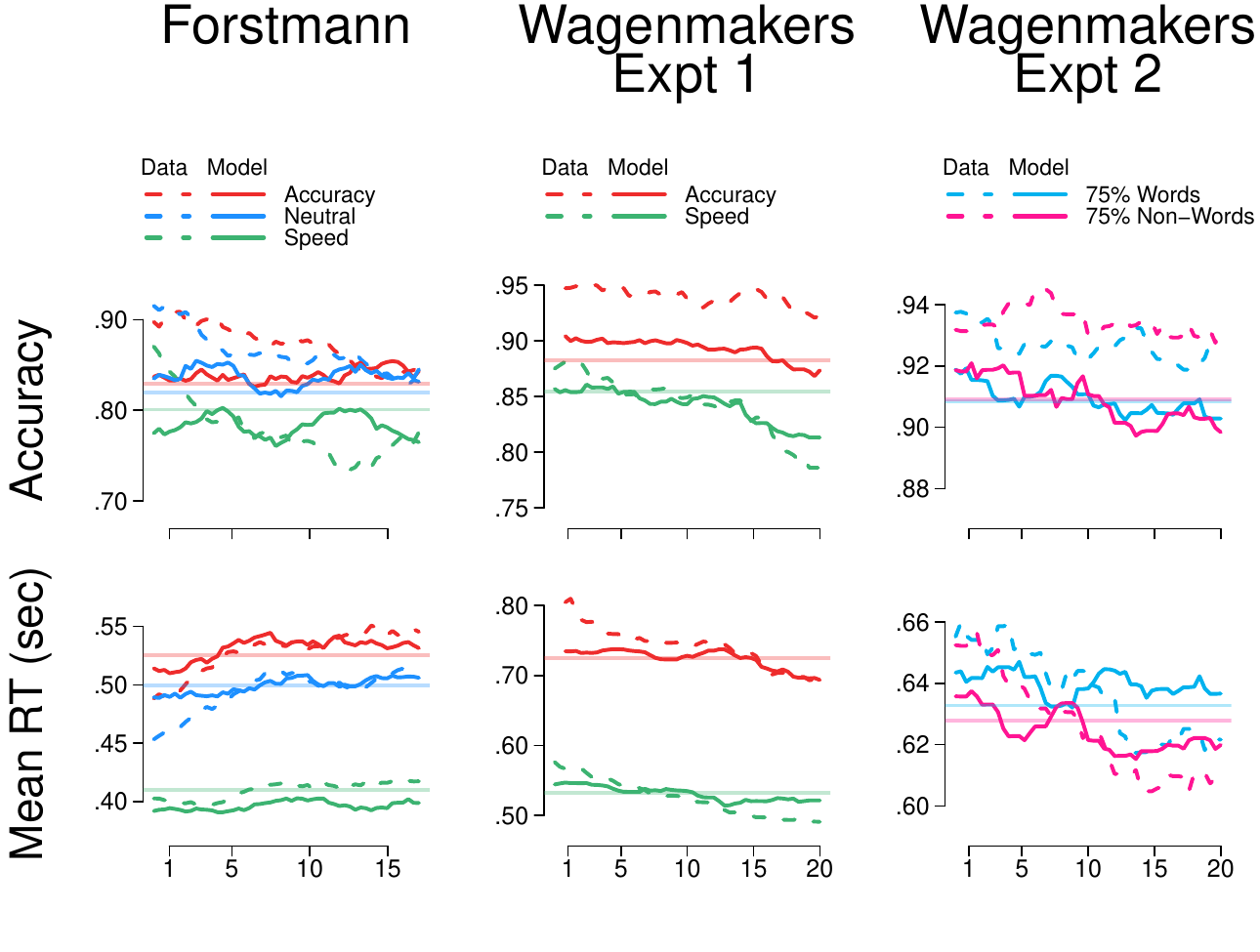}
\caption{\label{fig:goodness-of-fit} Data (dashed lines) from experiments reported by Forstmann et al. (2008) and Wagenmakers et al. (2008), shown in columns. The rows show how average accuracy and average RT change with time-on-task ($x$-axis, with kernel smoothing). Posterior predictive data from the Markov switching model are overlaid as solid lines, and from the static model as transparent lines. }
\end{figure}

To examine goodness of fit, we compare posterior predictive data generated from the Markov switching model against the observed data using the proportion of correct responses and the mean RT. Figure~\ref{fig:goodness-of-fit} shows these comparisons broken down by time-on-task (trials -- smoothed) and by the important experimental conditions, but averaged across subjects. The Markov switching model captures the fundamental patterns in the data. Changes in threshold settings capture the differences in RT between experimental conditions well. The much smaller changes in accuracy between conditions are slightly less well captured. The Markov switching model is quite tightly constrained in accommodating time-on-task effects. The model predicts those changes by predicting changes in the proportion of time spent in the two hidden states. This requires the model to predict matching changes in both accuracy and RT. For the most part, this is also observed in the data, e.g., the declining accuracy and RT for Experiment 1 of \cite{wagenmakers2008diffusion}, but not always, e.g., the first phase of \citeauthor{forstmann2008striatum}'s (\citeyear{forstmann2008striatum}) experiment has rapidly decreasing accuracy along with increasing RT. Of note, the model under-predicts average accuracy in the easiest conditions of Wagenmakers et al.'s experiments -- those conditions where observed accuracy was near ceiling. This limitation is imposed by the tightly constrained parameterization of the basic decision-making model in each state. That parameterization is also the same as used for the static model, which shows the same under-prediction of the high-accuracy observations, and was chosen to match previous investigations of these data.

\begin{figure}
\centering
\includegraphics[width=0.95\textwidth]{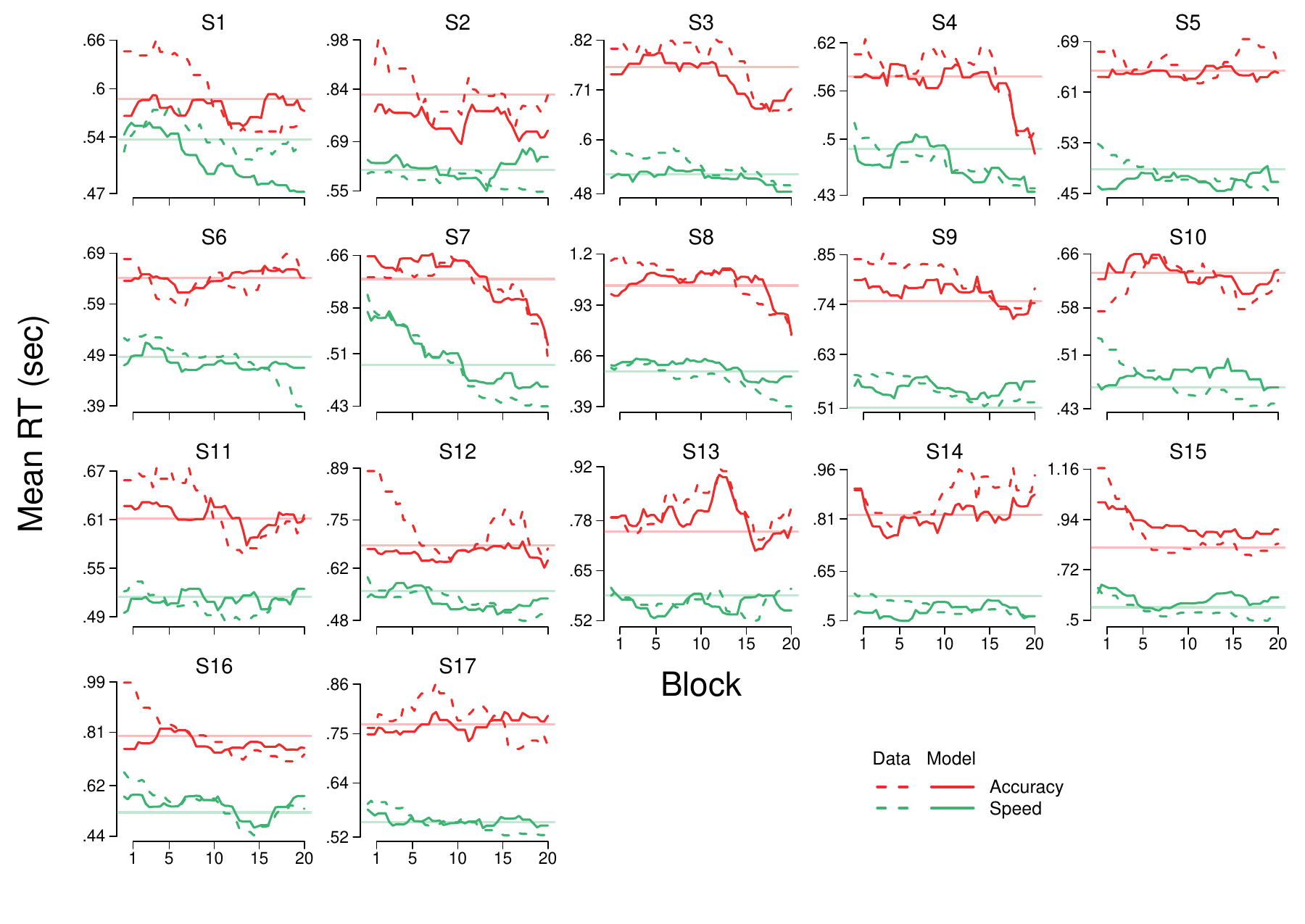}
\caption{\label{fig:goodness-of-fit-RT-LDT1} Mean RT (dashed lines) and model fit (solid lines for the Markov switching model; straight transparent lines for the static model) for Experiment 1 in Wagenmakers et al. (2008). Each panel shows a different subject. The red and green lines show performance in the accuracy and speed emphasis conditions, across time-on-task.}
\end{figure}

In addition to providing a good fit at the level of averaged data, the Markov switching model also provides a good fit to individual subjects' data, at least for most subjects. We generated plots similar to Figure~\ref{fig:goodness-of-fit}, but separately, for every individual subject. Figure~\ref{fig:goodness-of-fit-RT-LDT1} shows just one of six such plots; the fit to the RT data from \cite{wagenmakers2008diffusion}. Appendix \ref{appendix:additional-figures} contains the other five plots: accuracy fit for \citeauthor{wagenmakers2008diffusion}'s Experiment 1 data, and both accuracy and RT fits for \citeauthor{wagenmakers2008diffusion}'s Experiment 2 data and \citeauthor{forstmann2008striatum}'s data. Figure~\ref{fig:goodness-of-fit-RT-LDT1} shows that effects apparent in the group averaged fits are borne out in individual fits. The Markov switching model captures changes between conditions and changes with time-on-task for most individual subjects.

\begin{landscape}
\begin{table}
\caption{Group-level posterior mean parameter estimates for the two states of the Markov switching model, for each of the three experiments. Corresponding estimates for the static LBA model are also shown (in brackets) for comparison. Parameters are grouped by psychological processes: Caution ($b - \frac{A}{2}$); non-decision time ($\tau$); and drift rate ($v$). For drift rates in both of Wagenmakers et al.'s experiments the bracketed labels indicate the response accumulator (``W'' for ``word'' response and ``NW'' for ``nonword'' response) and stimulus condition (``hf'', ``lf'', ``vlf'' respectively for high frequency, low frequency and very low frequency words, and ``nw'' for nonwords). Statistically reliable differences between state 1 vs. state 2 are indicated by bold face. These correspond to less than 2.5\% overlap between posterior distributions across states.
}
\centering
\begin{tabular}{l||llll||llll||llll}
                        & \multicolumn{4}{l}{Forstmann}             & \multicolumn{4}{l}{Wagenmakers Exp. 1}      & \multicolumn{4}{l}{Wagenmakers Exp. 2}       \\
\hline
                   &     \textit{State}      & \textit{1}     & \textit{2} & \textit{(Static)}     &          & \textit{1}                           & \textit{2}   & \textit{(Static)}  &          & \textit{1}                           & \textit{2}  & \textit{(Static)}      \\
\hline
\hline
                        & Accuracy  & {\bf0.74} & {\bf0.92} & \textit{(0.98)}   & Accuracy & {\bf0.98 }                      & {\bf1.21}   & \textit{(1.37)} & (w,W)    & {\bf0.53}                        & {\bf0.83} & \textit{(0.96)} \\
          Caution              & Neutral   & {\bf0.68} & {\bf0.85}  & \textit{(0.91)} & Speed    & {\bf0.57}                       & {\bf0.75}  & \textit{(0.85)}& (w,NW)   & {\bf0.91}                        & {\bf1.22}  & \textit{(1.39)} \\
                        & Speed     & {\bf0.46} & {\bf0.56} & \textit{(0.63)}  &          &                             &   & & (nw,W)   & {\bf0.85}                         & {\bf1.14}  & \textit{(1.30)}   \\
                        &     & &      &       &        &          &                             &    & (nw,NW)  & {\bf0.54}                        & {\bf0.84}   & \textit{(0.96)} \\
\hline
$\tau$ &           & 0.23  & 0.21  & \textit{(0.18)} &          & 0.29                        & 0.20  & \textit{(0.15)} &          & 0.29                        & 0.23  & \textit{(0.18)}   \\
\hline
                        & Correct   & 1.16 & 1.32  & \textit{(1.40)}  & (hf,W)   & 3.39                        & 3.37  & \textit{(3.65)} & (hf,W)   & 3.27                        & 3.28  & \textit{(3.47)}\\
                        & Incorrect & 3.06 & 3.15   & \textit{(3.09)} & (lf,W)   & 2.71                        & 2.71  & \textit{(3.01)} & (lf,W)   & 2.55                        & 2.71  & \textit{(2.87)} \\
                        &        &    &       &       & (vlf,W)  & 2.34                        & 2.41   & \textit{(2.64)} & (vlf,W)  & 2.13                        & 2.35  & \textit{(2.51)} \\
          Drift              &   &          &       &        & (nw,W)   & {\bf0.37}                        & {\bf0.77}  & \textit{(0.67)} & (nw,W)   & 0.21                        & 0.22  & \textit{(0.52)} \\
          Rate              &    &        &       &       & (hf,NW)  & {\bf0.58}                        & {\bf1.07}  & \textit{(0.85)} & (hf,NW)  & 0.44                        & 0.42  & \textit{(0.64)} \\
                        &     &       &       &        & (lf,NW)  & {\bf0.68}                        & {\bf1.26}  & \textit{(1.15)} & (lf,NW)  & 0.67                        & 0.55  & \textit{(0.93)} \\
                        &       &     &       &       & (vlf,NW) & {\bf1.07}                        & {\bf1.57}  & \textit{(1.48)} & (vlf,NW) & 1.13                        & 0.94  & \textit{(1.31)} \\
                        &       &     &       &        & (nw,NW)  & 2.69                        & 2.67  & \textit{(2.96)}& (nw,NW)  & 2.60                         & 2.66  & \textit{(2.86)}
\end{tabular}
\label{tab:MSparams}
\end{table}

\end{landscape}

\subsection{Psychological Interpretation of the Markov Switching Model}

Table~\ref{tab:MSparams} reports the posterior mean estimates for the Markov switching model's parameters, for both hidden states and all three experiments. For the experiment reported by \cite{forstmann2008striatum}, and Experiment 2 of \cite{wagenmakers2008diffusion}, the model has identified two states which differ \emph{only} in the level of decision caution. For those experiments, the drift rate and non-decision time parameters are not reliably different between states; the posterior distributions for the difference between states includes zero for all those parameters. However, in both experiments, subjects adopt a more cautious speed-accuracy trade-off strategy in State \#2 than State \#1. It is common to operationalize decision caution as the average amount of evidence that must be accumulated to reach the decision threshold \citep{forstmann2008striatum,rae2014hare}, which is $b-\frac{A}{2}$. For the three conditions in \citeauthor{forstmann2008striatum}'s experiment, caution is reliably larger in State \#2 than State \#1, as indicated by the 95\% credible intervals for the difference in caution between states, which excluded zero in all three conditions: accuracy $(0.06,0.32)$; neutral $(0.06,0.29)$; speed $(0.03,0.19)$. For each of the four threshold estimates in \citeauthor{wagenmakers2008diffusion}'s Experiment 2, caution is reliably higher in State \#2 than State \#1: $w,W=(.15,.44)$; $w,NW=(.11,.50)$; $nw,W=(.12,.48)$; $nw,NW=(.15,.44)$. This is also true for the two estimates in \citeauthor{wagenmakers2008diffusion}'s Experiment 1: accuracy $(.05,.42)$; speed $(.07,.31)$.

The Markov switching model leads to different psychological interpretations for Experiment 1 of \cite{wagenmakers2008diffusion}. From those data, the model identifies hidden states which differ in both the level of caution adopted and the level of response inhibition engaged by the decision-maker. As for the other experiments, State \#1 implied more urgent decision-making than State \#2 -- less distance from start of evidence accumulation to threshold, on average. In addition, drift rate differences between states imply changes in the rate of evidence accumulation which are psychologically meaningful. The speed of evidence accumulation, i.e., drift rate, for accumulators which correspond to \emph{correct} response choices do \emph{not} differ between states. These are drift rates for the accumulator corresponding to the ``nonword'' response, for the trials when the stimulus is a nonword -- parameter ``nw,NW'' in Table \ref{tab:MSparams} -- and the drift rates for the accumulator corresponding to the ``word'' response for trials when the stimulus is a word of any frequency -- parameters ``hf,W'', ``lf,W'', and ``vlf,W'' in Table~\ref{tab:MSparams}. Instead, the differences between states are confined solely to the rate at which evidence accumulates for \emph{wrong} responses: drift rates for the ``word'' accumulator during nonword trials, and the ``nonword'' accumulator during word trials; parameters ``hf,NW'', ``lf,NW'', ``vlf,NW'', and ``nw,W'' in Table \ref{tab:MSparams}. Since these parameters represent the rate of accumulation of evidence in favor of incorrect responses,  higher values indicate poorer performance. In State \#2, the drift rates for incorrect choices are larger by between 0.40 and 0.59 units than in State \#1. These differences between states are statistically significant, posterior distributions for the differences have 95\% credible intervals that exclude zero: $nw,W=(.11,.73)$; $hf,NW=(.10,.91)$; $lf,NW=(.17,1.01)$; $vlf,NW=(.08,.94)$. To put the magnitude of the differences in context, the difference between the drift rates for the two accumulators in State \#1 is about 1.27 in the most difficult condition, with very low frequency word stimuli. This implies that between 30\% and 45\% of the subjects' sensitivity to the core decision task is lost in State \#2. The psychological implication of our analyses of Experiment 1 of \cite{wagenmakers2008diffusion} is that subjects switched between two cognitive states during performance. In one of those states, subjects are  much less effective at appropriately suppressing incorrect responses. This represents a weakening of executive functions related to inhibitory control \citep{karayanidis2009anticipatory,monsell2006can,monsell2003task}.

Table~\ref{tab:MSparams} also shows, in brackets, parameter estimates from static LBA fits to the three data sets. Naively, one might expect that parameters estimated from the static model might fall mid-way between the parameters estimated for the two hidden states of the Markov switching model. Table~\ref{tab:MSparams} shows that this is not the case. Instead, it appears that the standard approach of fitting a static model to data which are more complex introduces estimation bias. The estimated time taken by non-decision process (parameter $\tau$) is substantially smaller for the static model than for either state in the Markov switching model, in all three experiments. Given existing knowledge about the speed of simple detection tasks \citep[e.g.,][]{luce1986response} it seems likely that the slower estimates -- from the Markov switching model -- are closer to correct than the faster estimates from the static model. Similarly, estimated caution levels are larger from the static model than from the Markov switching model, in every condition, for all three experiments. For the most part, drift rate estimates are larger for the static model than for the Markov switching model, although there are exceptions in a small number of conditions (drift rates for the accumulator corresponding to the incorrect response, in Experiment 1 of Wagenmakers et al., 2008). 

\subsection{Longer-Term Dynamics}

\begin{figure}
\centering
\includegraphics[width=0.75\textwidth]{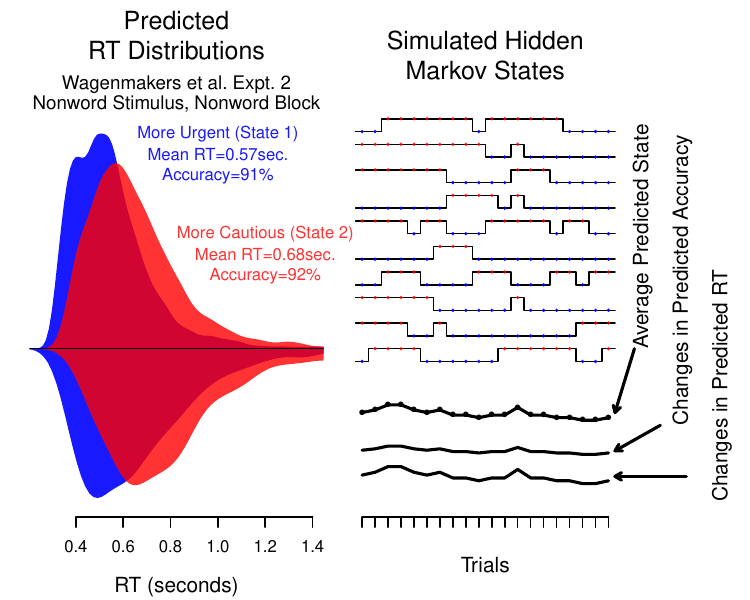}
\caption{\label{fig:MSLBAcartoon} Illustration of the Markov switching model, using estimates from Experiment 2 of \cite{wagenmakers2008diffusion}. The left panel shows the predicted RT distributions corresponding to the two states (different colors). These predictions are generated using the mean group-level posterior parameters, for one example condition (a non-word stimulus presented during a block with 75\% non-words). Distributions of correct and incorrect choices are shown above and below the horizontal axes, respectively. The two states correspond to more cautious vs. more urgent decision-making, leading to changes in RT but negligible effects on accuracy. The right panel shows posterior samples from the Markov process, switching between states (stepped lines with colored dots). The three solid lines below show the corresponding average predictions for the probability of being in each state, and subsequent accuracy and mean RT.}
\end{figure}

Figure~\ref{fig:MSLBAcartoon} illustrates the Markov switching model in detail, using the actual (group mean posterior) parameter estimates from Experiment 2 of \cite{wagenmakers2008diffusion}. The left panel shows the predicted response time and accuracy from State \#1 vs \#2, using different colors. The RT distributions above and below the horizontal centre line show predictions for correct and incorrect responses, respectively. Predictions for the observed data arise from mixtures of responses generated from the two states. The mixing is governed by the hidden Markov process illustrated in the right-hand half of the figure. There, the stepped lines with colored dots indicate sample trajectories of the Markov process, showing how the internal state switches randomly every few trials. The solid lines below the stepped traces illustrate how changes in state can lead, on average, to time-varying effects as observed in the data. 

\begin{figure}
\centering
\includegraphics[width=0.5\textwidth]{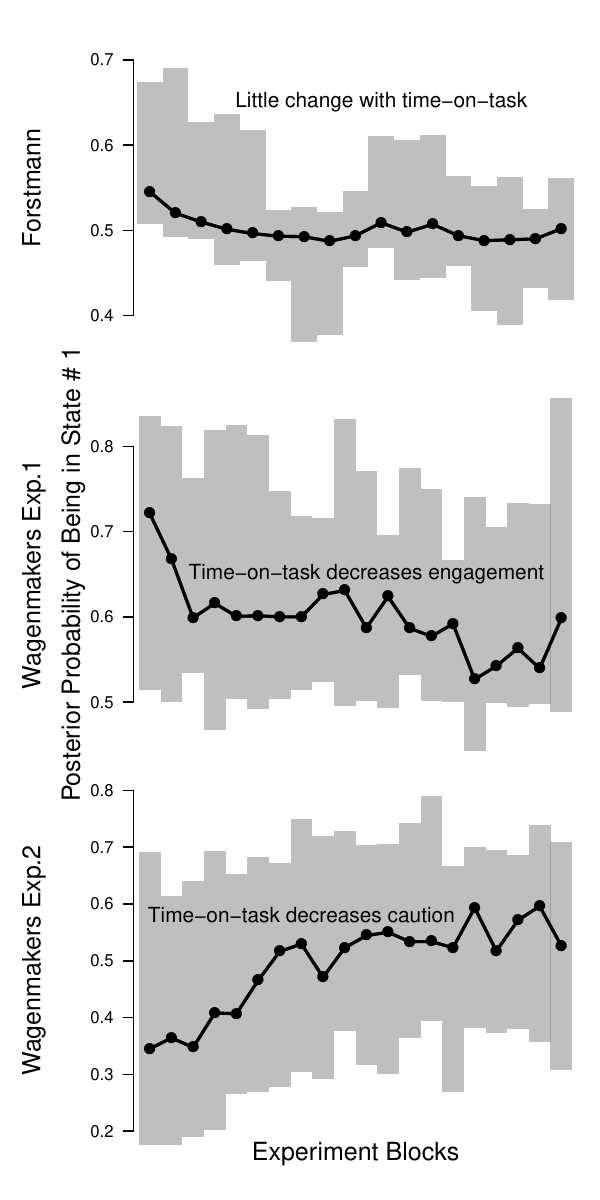}
\caption{\label{fig:stateprobs} Time-on-task effects are captured by the Markov switching model via changes in the posterior probability of the states. Over blocks of the experiments ($x$-axes), the posterior probability of decisions arising from State \#1 changes. The solid lines with dots show the median, across subjects, of this posterior probability. The shaded rectangles indicate between-subject variability, by showing the inter-quartile range for each block. For the two experiments reported by \cite{wagenmakers2008diffusion}, these results are consistent with a steady decrease in the care taken by subjects over time: for Experiment 1, this was caused by decreasing use of the more-engaged State \#1; for Experiment 2, this was caused by increasing use of the less-cautious State \#1. For the experiment of \cite{forstmann2008striatum}, there is an initial small change followed by a long period of stable performance.}
\end{figure}

Although the Markov process assumed a stationary prior, the time-on-task effects are captured by changes in the posterior distribution of state probabilities. Figure~\ref{fig:MSLBAcartoon} also illustrates this process. The three solid lines below the sample trajectories (stepped lines) show how the average time spent in each state (averaged across sample trajectories) changes, and how this subsequently changes the predicted mean RT and accuracy. Figure \ref{fig:stateprobs} shows these effects as changes in the probability of being in State \#1 over blocks, as well as the between-subject variability in state probabilities. Performance in \citeauthor{forstmann2008striatum}'s study initially shows a slight decrease in the probability of being in the less cautious State \#1, followed by a long period of stable performance, with approximately half of the data generated from each state. Time-on-task effects are larger in the two experiments reported by \citeauthor{wagenmakers2008diffusion}. In both cases, the model reveals an increasing tendency for subjects to adopt less thorough decision-making approaches over time, but with different psychological causes in the two experiments. For Wagenmakers et al.'s Experiment 1 (middle panel), the model identifies two states which differ in how engaged the subjects are. The probability of adopting the more engaged decision-making State \#1 decreases rapidly over the first three blocks, and continues to decline slowly thereafter. This reflects decreasing levels of both caution and inhibitory control over the course of the experiment. In Experiment 2 (lower panel), the model identifies an increasing reliance on State \#1 over blocks, corresponding to less cautious decision-making. In the first few blocks, the lower-caution State \#1 generates around one third of the data, but this flipped such that in the final few blocks that state generates nearly 60\% of the data.

\subsection{Shorter-Term Dynamics} 

\begin{figure}
\centering
\includegraphics[width=0.6\textwidth]{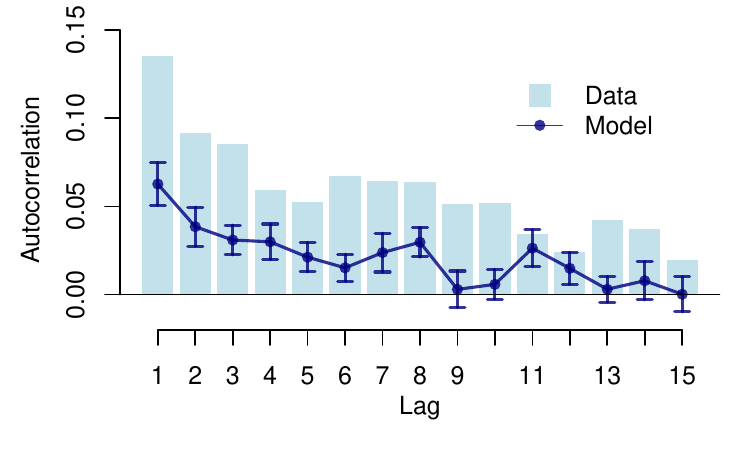}
\caption{\label{fig:acf} Autocorrelation of response times. Light blue bars show autocorrelation functions from data averaged over subjects from \cite{forstmann2008striatum}. Dark blue lines show the same from posterior predictive data generated by the Markov switching model (error bars show posterior standard deviations).}
\end{figure}

A ubiquitous feature of decision-making data -- indeed, behavioral data from many paradigms -- is correlation between successive observations. In response time data, this is often observed in autocorrelation functions which persist across several, or even dozens, of trials \citep[e.g.,][]{stewart2005absolute,stewart2009relative,brown2008integrated,vickers1998dynamic,vickers2000dynamic}. These autocorrelations are also observed in the data we investigated: Figure~\ref{fig:acf} shows the autocorrelation function for the data from \cite{forstmann2008striatum}, with persistent correlation between response times on decisions up to 15 trials apart (bars). An interesting consequence of the Markov switching model is its potential to provide a psychological explanation for the process which produces autocorrelation. Under the Markov switching model, nearby trials are more likely to be produced from the same hidden state than trials far apart in time, because every intervening trial adds another chance for a change in hidden state, leading to changed behavioral predictions. Figure~\ref{fig:acf} demonstrates that the Markov switching model reproduces the basic qualitative properties of autocorrelation in the data (dark blue line). This result is more impressive because the model is not constructed with the goal of explaining microstructure such as autocorrelation -- the autocorrelation is a natural consequence of the model structure. This is in contrast, for example, to theories such as SAMBA \citep{brown2008integrated} and PAGAN \citep{vickers1998dynamic} which ``build in'' autocorrelation via complicated additional mechanisms. The Markov switching model tends to underestimate the magnitude of autocorrelation, at most lags. One interpretation of this underprediction is that the Markov switching model provides an explanation for one source of autocorrelation in the data, but not all sources.

\section{Discussion}

Although human behavior is dynamic, analysts often treat data as if repeated observations are independent and identically distributed. Improvements gained through practice are often large, and are reliably observed across a wide range of behavior, from very simple counting tasks to complex tasks with multiple stages \citep{newell1981mechanisms,evans2018refining,wynton2017abrupt}. Time-on-task can also lead to decreases in performance, via fatigue or interference \citep{van2003mental,dorrian2007simulated,kahana2018variability} and also to changing periods of good and bad performance, caused by poor cognitive control and ``mind-wandering'' \citep[also called ``task unrelated thoughts'':][]{mittner2016neural,giambra1995laboratory}. In some cases, cognitive changes with time-on-task are a central focus of the research activity, and are directly investigated.

When off-the-shelf analysis tools such as ANOVA and other general linear models are applied, the data are routinely aggregated across time-on-task. This is not an inherent limitation of the analysis, but a convention with rare exceptions: for example, \cite{kahana2018variability} show how time-on-task effects including random changes, fatigue, and practice can all be described and measured using general linear models. Dynamic effects are often also ignored when more meaningful psychological theories are applied. Applications of process-based models in psychology most often assume an unchanging generative process across time, and identically distributed data across an experiment. There are dozens, or perhaps hundreds, of model-based analyses of the cognitive processes underlying simple decision-making using evidence accumulation models similar to the one used above. These include applied studies which reveal how experimental manipulations or person-based variables are associated with differences in decision-making due to changed caution, sensitivity, or other model parameters, while other studies focus on developing or elaborating the decision-making theories themselves, including extension; see \cite{ratcliff2016diffusion} for a review. 
Our work provides an approach for resolving discrepancies between the ubiquitous effects of time-on-task and the psychological theories which mostly ignore them. We extend a standard static model of decision-making, the linear ballistic accumulator, in psychologically-meaningful ways to account for time-varying behaviour. We explore extensions in which the model's parameters evolve with time on task. The most successful extension is a regime switching model which allows for decision-making behavior to be generated from one of two latent states, with different parameters. Behaviour switches between states according to a hidden Markov process, which represents a plausible process-level theory for the evolution of psychological states \citep[see also:][]{visser2009hidden,hamaker2010regime,busemeyer2009empirical}.


\subsection{Different Approaches to Time-Varying Effects}
Some earlier work takes into account time-varying effects in data, in different ways. For example, the PAGAN model of \cite{vickers1998dynamic} and the SAMBA model of \cite{brown2008integrated} extend evidence accumulation models to cover some time-varying effects. These models are based on accumulators racing to make decisions, but with elaborated elements describing how the properties of the accumulators change from decision-to-decision; e.g., due to meta-cognitive changes in confidence, or to facilitate repeated responses. These models are very interesting and have had substantial success in expanding understanding in their respective fields. However, incorporating time-varying effects incurs a substantial cost -- those models are intractable, and cannot be meaningfully estimated from the data. Their authors demonstrate that the models reproduce important patterns in the data with hand-tuned parameter settings, but this is very different from our approach. Our time-varying models allow accurate estimation of parameters from data. This puts them in the class of models that are used for ``cognitive psychometrics'', drawing psychologically-meaningful inferences from data by comparing model estimates between conditions and groups. Other work has extended reinforcement learning models to incorporate evidence accumulation processes. Most successful in this regard are the combinations of reinforcement learning models and diffusion models \citep[RL-diffusion:][]{pedersen2017drift,pedersen2020simultaneous,wiecki2013hddm}; although see also \cite{miletic2021new}. These models describe how the drift rate of a diffusion model can be linked to the output of a reinforcement learning model. The resultant combination can capture time-varying effects in experiments where those effects occur due to value learning or reinforcement and reward. The more general application of RL-diffusion is limited by both scientific considerations related to changes caused by any effects other than reinforcement learning, and statistical considerations. For example, the three data sets analyzed above all include within-subject manipulations (like almost all psychological experiments) and time-varying effects not caused by reinforcement. Both of these are not easy to manage in the RL-diffusion framework developed of \citeauthor{pedersen2020simultaneous}. 

Other earlier investigations of time-varying effects make an opposite compromise, by replacing detailed psychological models of the basic psychological process with simplified, descriptive, accounts. For example, \cite{craigmile2010autocorrelated} studied time varying effects in RT data, and developed sophisticated model estimation methods; see also \cite{peruggia2002car}. These authors treat the decision-making process descriptively, for example using a Weibull distribution for response times, which limits the inferences that can be drawn about psychological processing. \cite{kahana2018variability} adopted a similar approach for studying recall performance over many sessions of a memory study. \citeauthor{kahana2018variability} used a general linear model with mixed effects to tease apart contributions due to practice, fatigue, the memory materials, and other variables. Although in very different domains, the approaches of \citeauthor{peruggia2002car}, \citeauthor{craigmile2010autocorrelated}, and \citeauthor{kahana2018variability} share a similar methodological approach based on descriptive statistical models of performance. This approach provides statistical convenience but limits the ability to investigate underlying psychological processes. For example, the analyses used by \citeauthor{peruggia2002car} do not separate out different psychological components which contribute to changes in RT, such as the effects of caution vs. processing speed, and the analyses used by \citeauthor{kahana2018variability} do not separate out effects on recall performance due to response strategies vs. memory strength. 
Our approach is an important advance that uses a psychological process model of the underlying phenomenon, like \cite{vickers1998dynamic} and \cite{brown2008integrated}, but it also supports accurate parameter estimation, and practical use, similarly to \cite{craigmile2010autocorrelated}, \cite{kahana2018variability}, and \cite{peruggia2002car}.

\subsection{Future Directions and Conclusions}
Of the three accounts we investigate for time-varying effects, only the Markov switching model represents a psychologically meaningful process-level explanation \citep[for review, see][]{visser2011seven}. This makes it heartening that the Markov switching model is also preferred by statistical model selection with marginal likelihood. The switching model adopts the plausible assumption that observed behavior is actually a mixture over states; we assume two states, but this is not a strong commitment. The mixing process itself is further explained by a hidden Markov process: behavior is always generated from just one of the two states, and at each moment there is a fixed probability of switching into the other state. This theory entails interesting links to research on ``Type 1'' vs. ``Type 2'' decision-making \citep{newell2014unconscious,kahneman2011thinking} as well as mind-wandering \citep{mittner2014when,mittner2016neural}.

Our models and estimation methods may be useful in future applications to decision-making data. One way to use them is to screen data for the presence of important time-on-task effects. Identifying these effects is usually harder than plotting the data, because the effects can disappear in averages, and be nonlinear in individuals. Model comparison between the static and time-varying versions of the LBA can reveal the presence of time-varying effects. Analysis of the parameter estimates can help to interpret the nature of time-on-task effects, and identify the underlying psychological causes. Our modeling approach is likely to be useful beyond the specific application to the LBA model, and beyond decision-making research more generally. The central idea involves incorporating statistically tractable models for time series as descriptions of the changes in model parameters for individual subjects with time-on-task. This approach presents some estimation challenges, which are surmountable for many psychological theories, given recent advances in statistical and computational approaches. We hope that our approach  is used in two ways: as a way to routinely model time-on-task effects in simple decision-making; and, more generally, to extend quantitative theories of cognition to investigate interesting effects such as those of practice, learning, and fatigue.

\newpage 

\bibliography{uoncoglabshort}

\newpage

\appendix

\section{\textbf{Additional Figures}\label{appendix:additional-figures}}

\begin{figure}[H]
\centering
\includegraphics[width=0.9\textwidth]{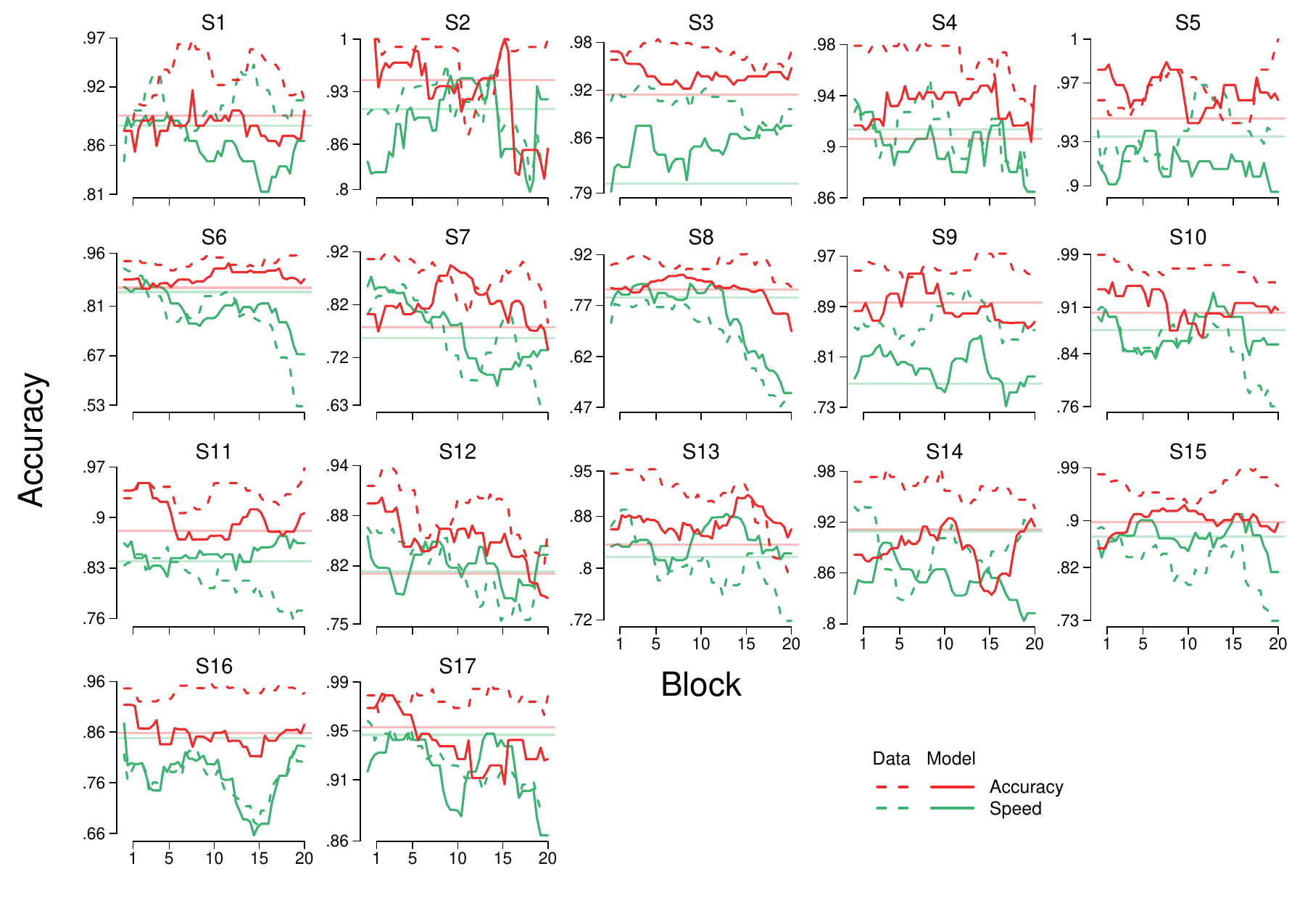}
\caption{\label{fig:goodness-of-fit-acc-LDT1} Decision accuracy from Experiment 1 of Wagenmakers et al. (2008). Each panel shows a different individual subject. The dashed lines in red and green show performance in the accuracy and speed emphasis conditions, across time-on-task (blocks; $x$-axis). The solid lines show the posterior predictive data from the Markov switching model. The straight transparent lines show the posterior predictive data of the static LBA.}
\end{figure}

\begin{figure}[H]
\centering
\includegraphics[width=0.9\textwidth]{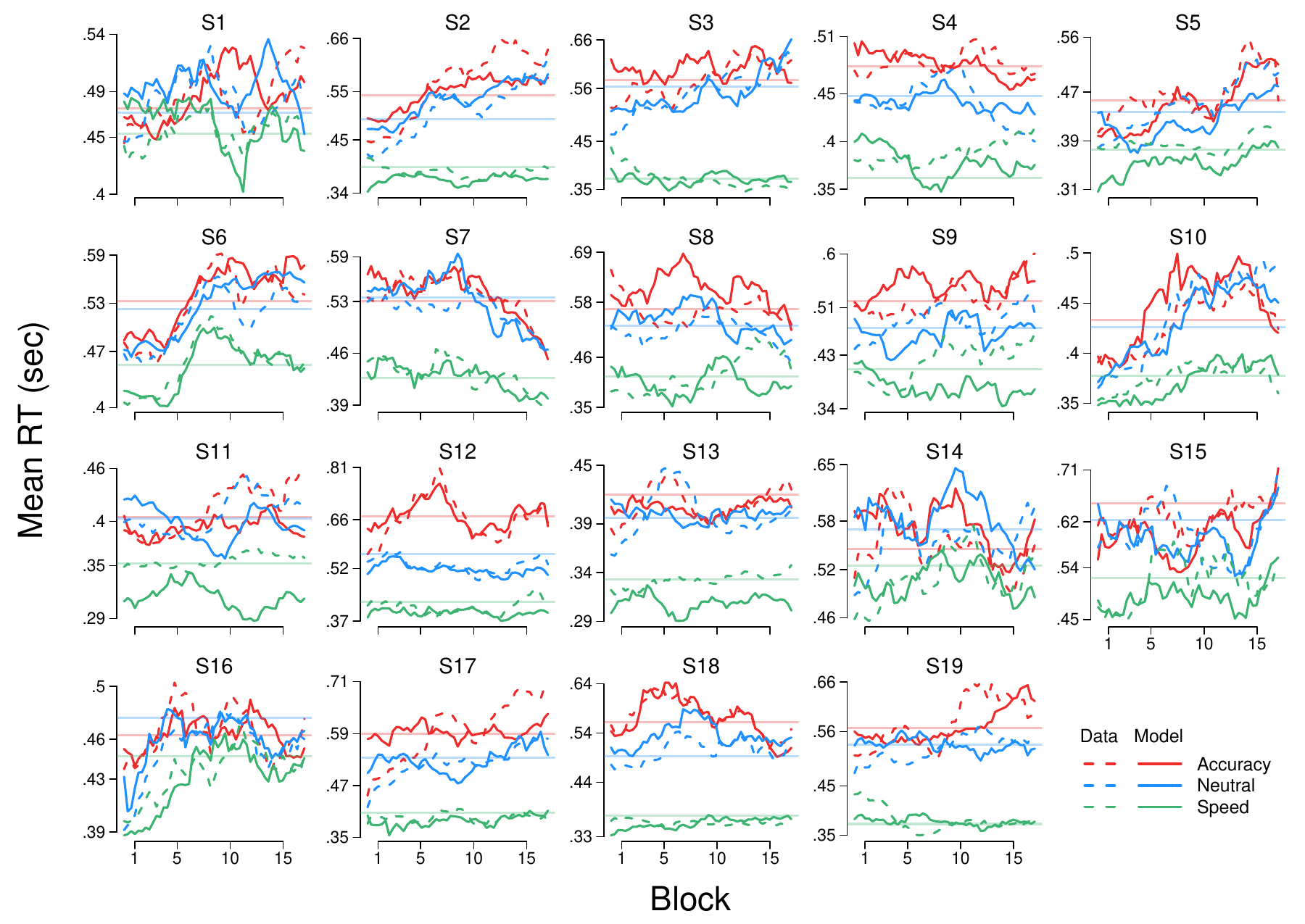}
\caption{\label{fig:goodness-of-fit-RT-forstmann} Mean RT from Forstmann et al.'s (2008) experiment. Each panel shows a different individual subject. The dashed lines in red, blue and green show performance in the accuracy, neutral and speed emphasis conditions, across time-on-task (blocks; $x$-axis).  The solid lines show the posterior predictive data from the Markov switching model. The straight transparent lines show the posterior predictive data of the static LBA.}
\end{figure}

\begin{figure}[H]
\centering
\includegraphics[width=0.9\textwidth]{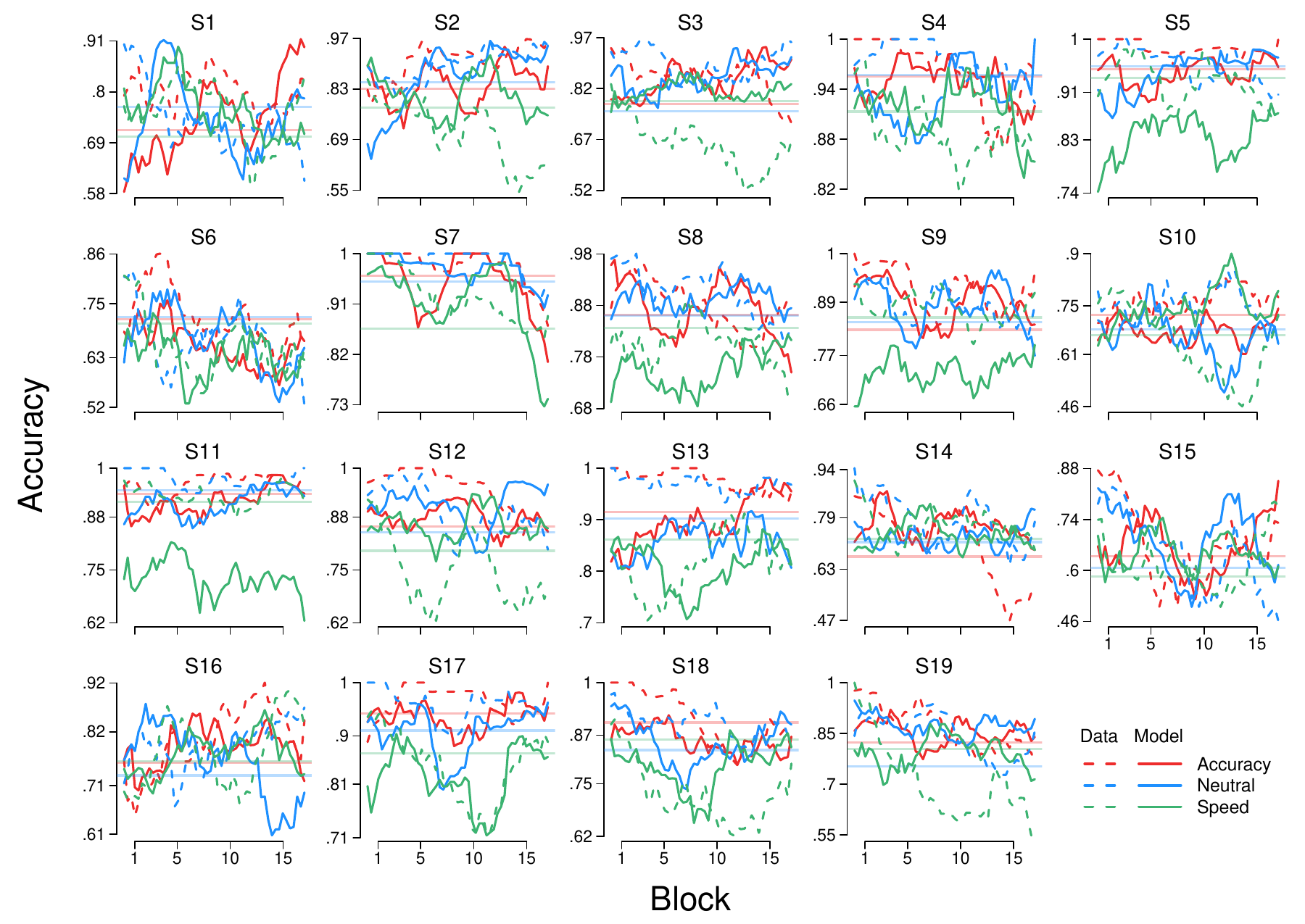}
\caption{\label{fig:goodness-of-fit-acc-forstmann} Decision accuracy from \citeauthor{forstmann2008striatum}'s \citeyear{forstmann2008striatum} experiment. Each panel shows a different individual subject. The dashed lines in red, blue and green show performance in the accuracy, neutral and speed emphasis conditions, across time-on-task (blocks; $x$-axis). The solid lines show the posterior predictive data from the Markov switching model. The straight transparent lines show the posterior predictive data of the static LBA.}
\end{figure}

\begin{figure}[H]
\centering
\includegraphics[width=0.9\textwidth]{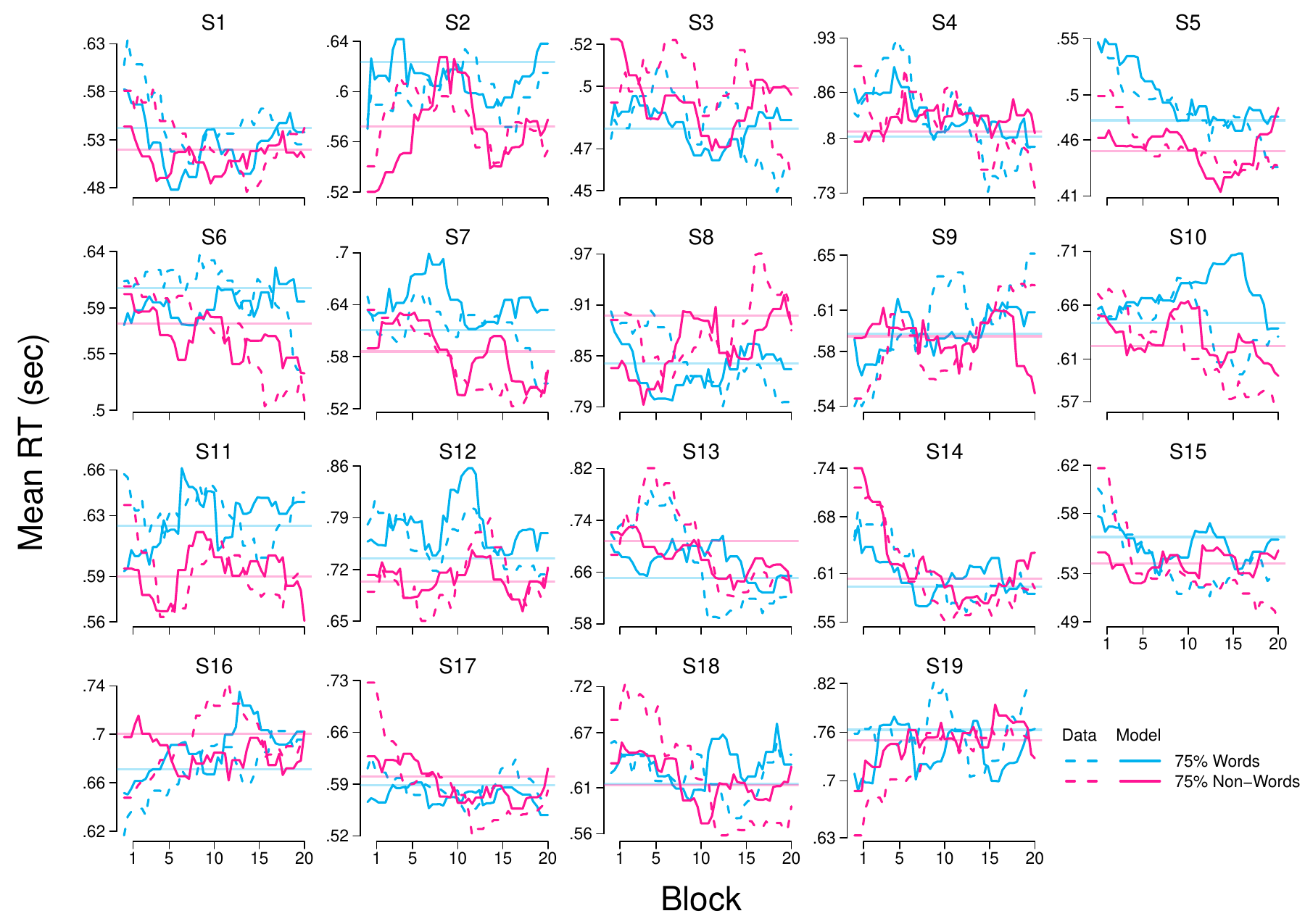}
\caption{\label{fig:goodness-of-fit-RT-LDT2} Mean RT from Experiment 2 of \cite{wagenmakers2008diffusion}. Each panel shows a different individual subject. The dashed lines in blue and pink show performance in the majority word and majority non-word conditions, across time-on-task (blocks; $x$-axis). The solid lines show the posterior predictive data from the Markov switching model. The straight transparent lines show the posterior predictive data of the static LBA.}
\end{figure}

\begin{figure}[H]
\centering
\includegraphics[width=0.9\textwidth]{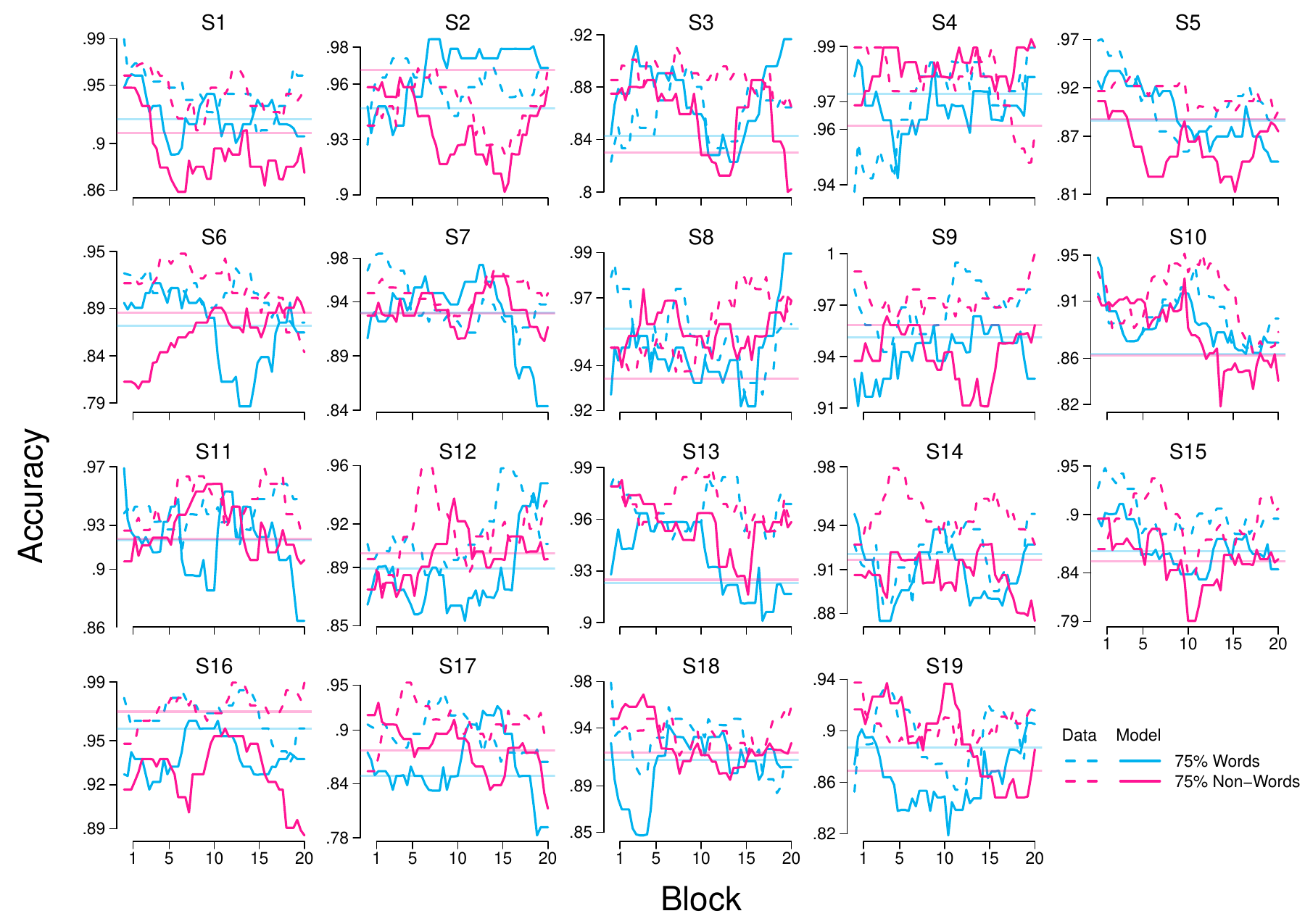}
\caption{Decision accuracy from Experiment 2 of \cite{wagenmakers2008diffusion}. Each panel shows a different individual subject. The dashed lines in blue and pink show performance in the majority word and majority non-word conditions, across time-on-task (blocks; $x$-axis). The solid lines show the posterior predictive data from the Markov switching model. The straight transparent lines show the posterior predictive data of the static LBA.}
\label{fig:goodness-of-fit-acc-LDT2} 
\end{figure}

Figures \ref{fig:goodness-of-fit-acc-LDT1} to  \ref{fig:goodness-of-fit-acc-LDT2} show additional results for the goodness of fit of the Markov switching model and the static LBA model to individual subjects: decision accuracy for Experiment 1 of \cite{wagenmakers2008diffusion}, and mean RT and decision accuracy for \cite{forstmann2008striatum} and Experiment 2 of \cite{wagenmakers2008diffusion}.

\end{document}